\documentclass[prb,preprint,amsmath,amssymb,apl]{revtex4-1}
\usepackage{amsbsy}
\usepackage{amssymb}
\usepackage{graphicx}

\makeatletter

\usepackage{siunitx}
\begin{document}

\title{P-type $\delta$-doping with Diborane on Si(001) for STM Based Dopant
Device Fabrication}
\author{Tom\'a\v{s} \v{S}kere\v{n}}
\email{t.skeren@gmail.com}
\author{Sigrun K\"oster}
\affiliation{IBM Research \textendash{} Zürich, Saümerstrasse 4, 8803 Rüschlikon,
Switzerland} 
\author{Bastien Douhard}
\affiliation{IMEC, Kapeldreef 75, 3001 Heverlee, Belgium}
\author{Claudia Fleischmann}
\affiliation{IMEC, Kapeldreef 75, 3001 Heverlee, Belgium}
\author{Andreas Fuhrer}
\email{afu@zurich.ibm.com}
\affiliation{IBM Research \textendash{} Zürich, Saümerstrasse 4, 8803 Rüschlikon,
Switzerland}


\begin{abstract}
Hydrogen resist lithography using the tip of a scanning tunneling
microscope (STM) is employed for patterning p-type nanostructures
in silicon. For this, the carrier density and mobility of boron $\delta$-layers,
fabricated by gas-phase doping, are characterized with low-temperature
transport experiments. Sheet resistivities as low as $300\thinspace\Omega$
are found. Adsorption, incorporation and surface diffusion of the
dopants are investigated by STM imaging and result in an upper bound
of 2\,nm for the lithographic resolution which is also corroborated
by fabricating a 7.5\,nm wide p-type nanowire and measuring its electrical
properties. Finally, to demonstrate the feasibility of bipolar dopant
device fabrication with this technique, we prepared a 100\,nm wide
pn junction and show that its electrical behavior is similar to that
of an Esaki diode. 
\end{abstract}
\maketitle

\section{Introduction}

Definition of atomically sharp dopant profiles is a challenge in device
fabrication as scaling of electronic devices approaches the few nanometer
range~\cite{01FrankAA,08HoAA,14ColombeauAA,14DuffyAA}. Limits in
achievable doping densities and unwanted diffusion of the dopants
into the channel, impact both access resistance and variability of
today's semiconductor devices. On the other hand, single dopants have
become a resource as artificial atoms, embedded in near perfect cristaline semiconductors  sometimes referred to
semiconductor vacuum. They are used e.g. for qubits~\cite{14MuhonenAA}
or charge pumps~\cite{13RocheAA}. Hydrogen resist lithography using
an STM has achieved both atomically sharp dopant profiles~\cite{03SchofieldAA}
and deterministic patterning of single dopant devices~\cite{12FuechsleAA}.

Hydrogen termination of the Si(001) surface with a mono hydride strongly
suppresses the surface reactivity. Subsequent local desorption of
the hydrogen using the tip of an STM creates local reaction sites
(dangling bonds) where gas phase dopant precursors attach and can
be incorporated into substitutional silicon surface sites. For phosphine
this process is well established and low temperature incorporation
is possible without significant diffusion of the dopants (for a review
see Ref.\,\cite{05SimmonsAA}). However, degenerately doped p-type
acceptor devices have not been explored so far with this technique.
Patterning of acceptor structures at the nanoscale is crucial in terms
of the implications for fabrication of p-type field effect transistors,
but also to explore e.g. STM fabrication of bipolar planar dopant
devices or to exploit electric field tunability of acceptor spin states
for spin manipulation~\cite{15HeijdenAA}.

In the present work, we investigate gas-phase $\delta$-doping with diborane ($\mathrm{B_{2}H_{6}}$)
for p-type acceptor device fabrication with an STM, and find resistivities
that compare well with those of phosphorus $\delta$-doped layers
and donor devices~\cite{05SimmonsAA}. While full activation of boron
requires higher temperatures than for phosphorus, we show that patterning
of acceptor devices with hydrogen resist lithography is possible,
and blurring of the patterned device after annealing can be limited
to a few nanometers. We employ the p-type doping technique to fabricate an
acceptor nanowire and, in combination with n-type doping, a bipolar pn-junction. 

\section{Results and discussion}

In order to establish a method for p-type gas phase doping with diborane
for STM-based dopant device fabrication several
fundamental requirements need to be fulfilled. First,  conditions for
the reaction of diborane with the Si surface need to be established such that the surface is
saturated with the dopant precursor but avoiding unintentional
surface passivation with hydrogen gas, present in the precursor
as a carrier gas. Second, an important prerequisite for hydrogen
resist lithography is the selective adsorption of diborane to areas
of the Si:H surface which have been depassivated by the STM tip. Finally,
it is important to confirm that the boron $\delta$-layer can be encapsulated
by overgrowth with a Si layer without segregation, preserving its
atomic sharpness. The validity of these conditions is investigated
in the first part of the paper. 

In the second part we present electrical transport
properties of boron doped $\delta$-layers and their dependence
on diborane dose and activation temperature. 

In the last part we detail the fabrication procedure and electrical measurements
of two dopant devices which involve selectively diborane doped p-type areas. The
first one is a 7.5\,nm wide p-type nanowire and the second one is
a 100\,nm wide pn-junction which combines diborane p-type doping
with previously established n-type doping based on phosphine. 

\subsection{p-type $\delta$-doping}

\begin{figure}[b!]
\centering{}\includegraphics[width=130mm]{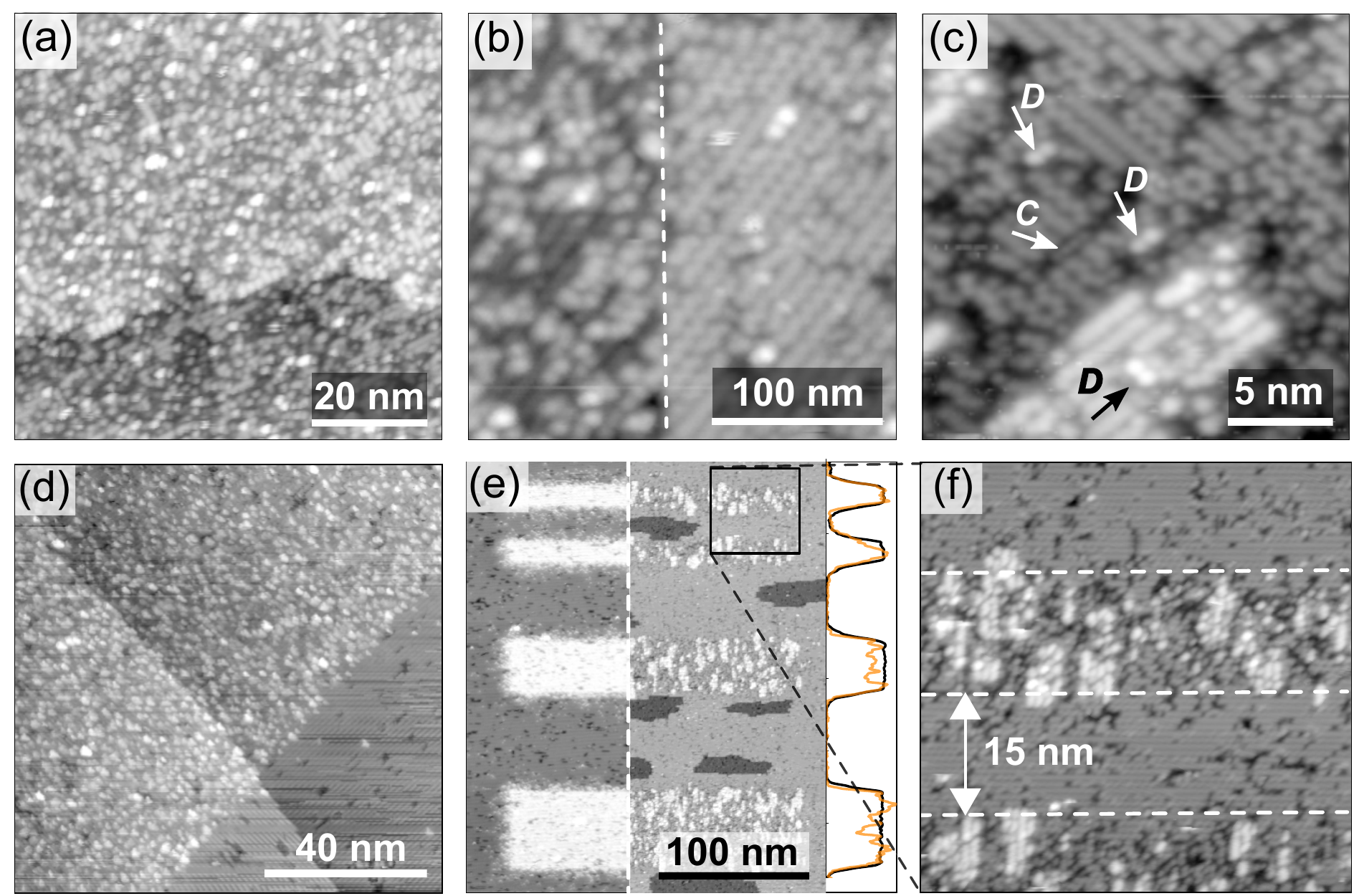} \caption{\label{fig:1_merge}STM ivestigation of adsorption and surface reactions
of diborane with with the Si(001) surface. (a) Si(001) surface prepared
by flash annealing in UHV and exposed to 3\,L of diborane (30\% in
H$_{2}$) at 200$\,^{\circ}$C. (b) Si surface exposed to 230\,L
of pure H$_{2}$ under similar conditions as in (a). The right half
was subsequently desorbed using the STM. (c) The surface shown in
(a) after annealing at 510$\,^{\circ}$C for 5\,min. \emph{C} and
\emph{D} denote features related to boron in the silicon surface.
(d) Edge of an STM-desorbed patch after exposure to 90\,L of diborane
at 180$\,^{\circ}$C. (e) Desorbed stripes after exposure to 90\,L
of diborane at 180$\,^{\circ}$C and a subsequent incorporation anneal
for 1\,min at 510\,$\,^{\circ}$C. To the right a scaled averaged line-cut of the stripes before (black) and after (brown) incorporation is shown. (f) Higher resolution image
of the narrowest stripes in (e).}
\end{figure}

In a first step, we use room temperature STM imaging to investigate
the sticking and surface reaction of diborane on Si(001). Previous
work found that saturation dosing of the Si(001) surface requires
doses of up to 1000\,L at room temperature~\cite{wang1996a,yu1986}.
However, heating of the silicon substrate was shown to improve sticking
of the diborane molecule due to improved surface dissociation. We
want to limit temperatures during dosing to below 450$\,^{\circ}$C
such that patterns written into the hydrogen resist remain intact~\cite{90OuraAA}.
Furthermore, while molecular hydrogen (the carrier gas) does not stick
to the silicon surface at room temperature, dissociative adsorption
of $\mathrm{H_{2}}$ on the silicon surface is known to increase with
temperature~\cite{durr2002}. This led us to select a sample temperature
of 200$\,^{\circ}$C during diborane dosing which ensures that structures
patterned using STM lithography stay intact and have acceptable doping
densities. 

Figure \ref{fig:1_merge}a shows a filled state STM image of a bare
silicon surface after exposure to 3\,L of diborane at 200$\,^{\circ}$C
(see Methods for details about dose interpretation). The image spans
two atomic silicon steps on which some residual silicon dimer rows
can be identified. We observe three distinct levels in the topography of each terrace 
where the residual dimer rows have an intermediate height. Furthermore,
there are both darker depressions and brighter protrusions. The dark
areas can be identified as the parts of the silicon surface that are
hydrogen terminated. Here, both the carrier gas $\mathrm{H_{2}}$
and the diborane molecule $\mathrm{B_{2}H_{6}}$ are potential sources
of hydrogen. Furthermore, it was previously suggested that some boron
hydride fragments may appear similar to a hydrogenated silicon dimer~\cite{wang1995}.

As with phosphine~\cite{06WilsonAA}, the maximum achievable boron
density through gas phase doping is expected to be limited by unintentional
surface passivation with hydrogen present in $\mathrm{B_{2}H_{6}}$~\cite{wang1996a}.
However, we also need to clarify the effect of the $\mathrm{H_{2}}$
carrier gas on this process~\cite{durr2002}. The surface in Figure~\ref{fig:1_merge}b
was exposed to 230\,L of pure molecular hydrogen at 220$\,^{\circ}$C.
We find that about half of the surface is covered by hydrogen through
surface dissociation, as shown in the left part of the STM image.
On the right side of the image hydrogen was subsequently desorbed
using the STM tip. This suggests that only a small fraction of the
hydrogen in Figure \ref{fig:1_merge}a, with a more than 20 times
smaller dose, originates from the $\mathrm{H_{2}}$ carrier gas. Our
choice of 200$\,^{\circ}$C as a dosing temperature therefore is a
good trade-off between the surface reactivities of $\mathrm{B_{2}H_{6}}$
and $\mathrm{H_{2}}$. After dosing we further heat the sample to
incorporate the boron into the silicon surface. Thermal decomposition
studies indicate that only hydrogen desorbs into the gas-phase while
boron remains on the surface and incorporates into substitutional
sites at temperatures above 400$\,^{\circ}$C~\cite{yu1986}. 

Figure \ref{fig:1_merge}c shows an image of the surface in \ref{fig:1_merge}a
after annealing to 510$\,^{\circ}$C for 5\,min. Here, we can identify
several features which have been attributed to boron atoms in the
surface, namely a double dot protrusion marked \textit{D} and an arrangement
of indented atoms marked \textit{C}. Both have been discussed before
for samples annealed at similar temperatures~\cite{wang1995,wang1996a,liu2008}.
While the \textit{C} feature is not easy to identify, the \textit{D}
feature seems to be typical for substitutional boron on Si(001) replacing
one of the two Si surface dimer atoms at low boron density~\cite{liu2008}.
In addition to this, there are still several depressions which we
attribute to strain induced defects in the silicon surface. These
STM images show that $\mathrm{B_{2}H_{6}}$ adsorbs on the silicon
surface at 200$\,^{\circ}$C and boron is at least partially incorporated
into the silicon crystal after further annealing. More precise quantification
of the effect of the incorporation anneal is obtained from transport
measurements on boron $\delta$-layers in the next section. 

In the following we now establish selectivity of the hydrogen resist
towards diborane dosing and investigate the persistence of the structures
during the annealing process. For this, we pattern the hydrogen terminated
Si surface using the STM tip and study the evolution of these structures
after dosing and annealing. Figure \ref{fig:1_merge}d shows adsorption
of the diborane on an STM patterned surface (upper left part has been
desorbed). After dosing the surface with 90\,L of diborane, the border
of a patterned patch remains sharply defined on the nanometer scale.
While a high coverage of bright protrusions is visible inside the
patch, the mono-hydride surface outside (lower right corner) remains
practically unaltered. 

The STM defined pattern remains stable even after annealing at temperatures
$T_{\mathrm{A}}$ high enough to desorb the hydrogen resist. Figure
\ref{fig:1_merge}e shows a pattern of 4 STM defined patches dosed
with 90\,L of diborane at \SI{180}{\celsius} after 1\,min anneal
to $T_{\mathrm{A}}=510\,^{\circ}$C. Even though clear changes in
surface morphology occur, the stripes are still well resolved. The
dark islands represent the next Si layer below the surface. They are
visible everywhere on the sample surface and are most likely caused
by etching during the removal of hydrogen. In the regions exposed
to diborane we find bright agglomerates which show dimer row formation
as can be seen in the zoom in Figure \ref{fig:1_merge}f. We believe
this to be ejected silicon atoms that have exchanged place with incorporated
boron and start to form islands due to diffusion. While it is difficult
to clearly identify the boron, we assume that the incorporated boron
atoms move no further than the ejected silicon on the surface. This
means that even at $T_{\mathrm{A}}=510\,^{\circ}$C, the resolution
of p-type dopant device patterning is expected to be significantly
better than 2\,nm. The two line-traces in \ref{fig:1_merge}e are horizontally averaged line cuts (arbitrarily scaled in z) of the four patches before and after annealing. The good overlap of the two traces further supports the high resolution of the pattern transfer.

$\delta$-doping and subsequent encapsulation by Si overgrowth can
be accompanied by another complication which has also been encountered
in the case of phosphine $\delta$-doping. During the overgrowth,
dopant segregation can lead to blurring of the $\delta$-doping
profile and, ultimately, deterioration of the inplane resolution \cite{Goh2004}.
In the case of phosphorus doping, this problem has been addressed
by using a so-called locking layer i.e. the first few monolayers of the
Si encapsulation layer are grown at lower temperature to suppress
segregation while the rest of the overgrowth is performed at an elevated
temperature in order to preserve the crystalline quality of the capping layer
and decrease the number of defects \cite{Keizer2015}. 

\begin{figure}[t!]
\centering{}\includegraphics{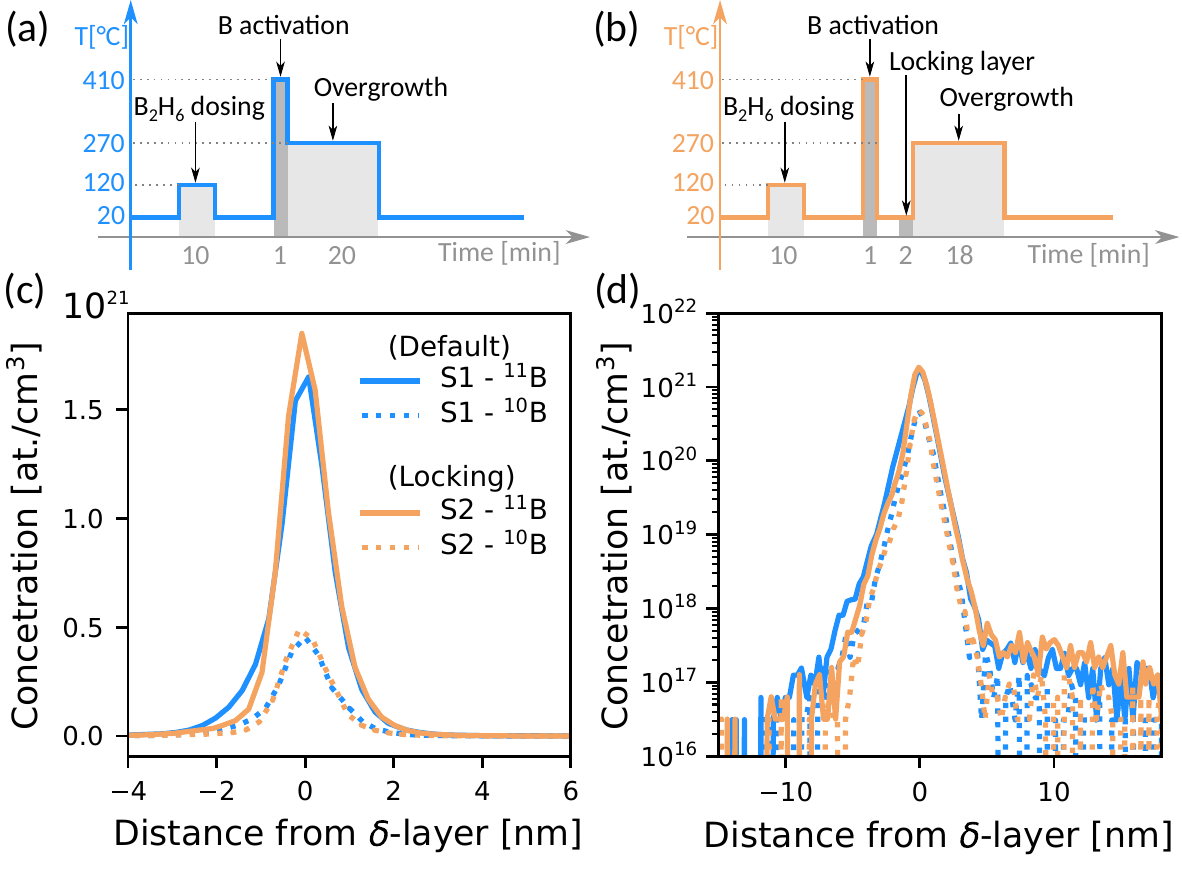}\caption{\label{fig_SIMS}Sample preparation and SIMS measurements of boron doped $\delta$-layers.
(a) Preparation procedure of sample S1 (default):  after dosing with diborane and a 1\,min activation anneal at \SI{410}{\celsius}
the $\delta$-layer is immediately overgrown with 20\,nm of intrinsic silicon
at a temperature of \SI{275}{\celsius}. (b) In
sample S2 (locking-layer) the sample is first cooled to room temperature after the activation anneal. A 2\,nm thick layer of Si is grown before raising the sample temperature to \SI{275}{\celsius} and growing the rest of the Si layer. (c, d) SIMS concentration profiles of the two samples: S1 (blue) and S2 (brown). Solid lines are for $^{11}$B dashed lines for $^{10}$B.}
\end{figure}

In boron $\delta$-doped layers we find that segregation is less of a concern. Secondary ion mass spectroscopy (SIMS) measurements were performed on two samples: S1 (default preparation as shown in Fig. \ref{fig_SIMS}a ) and S2 (includes locking-layer as shown in Fig. \ref{fig_SIMS}b ).

SIMS traces for $\mathrm{^{11}B}$ (solid lines) and $\mathrm{^{10}B}$ (dashed lines) are shown in Figs. \ref{fig_SIMS}c (linear scale) and \ref{fig_SIMS}d (logarithmic scale) with the correct isotopic ratio and for both samples. The consistent behavior of the two isotopes for each sample is a confirmation that the SIMS data is not influenced by other atomic species. 

For both samples the leading edge (left side) of the SIMS profile has a slightly smaller slope indicating that there is some boron segregation during overgrowth. This is stronger for the trace of sample S1 (blue trace) which exhibits a shoulder towards the surface. This means that 
the locking layer can indeed improve the sharpness of the boron $\delta$-doping profile, however, the measured characteristics are very similar for both samples and, compared to phosphorus doping,
segregation of the dopants does not represent a substantial
issue. Furthermore, a FWHM of 1.4\,nm for the $\delta$-layer thickness is already close to the theoretical
resolution limit of the SIMS technique and gives an
upper limit of the real thickness. These SIMS measurements therefore show that dopant segregation is
negligible when using boron as a dopant and the use of a locking layer is not necessary. This can be seen as a clear advantage over phosphorus since the required low temperature growth is usually associated with the formation of defects and lower overgrowth Si layer quality \cite{Keizer2015}. 

\subsection{Transport measurements on p-type $\boldsymbol{\delta}$-layers }

\begin{figure}[t!]
\centering{}\includegraphics{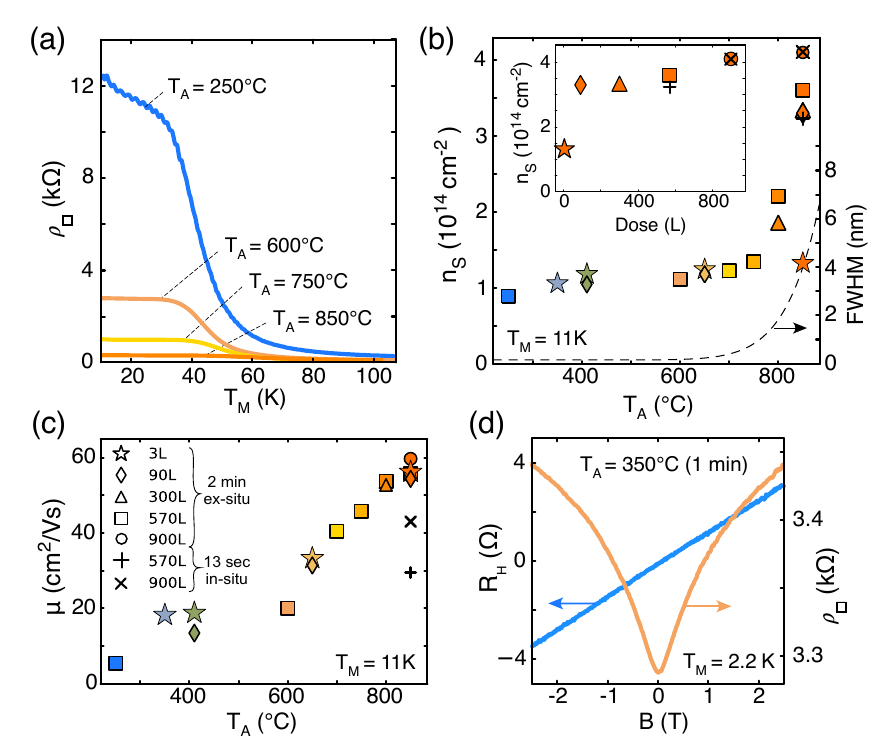}
\caption{\label{fig:2} (a) Resistivity of a 570\,L $\delta$-layer sample
as a function of measurement temperature $T_{\mathrm{M}}$ annealed
ex-situ at $T_{\mathrm{A}}$ for 2\,min. (b) Sheet density and (c)
mobility as function of annealing temperature $T_{\mathrm{A}}$. The
inset in (b) shows the weak dependence of $n_{\mathrm{S}}$ on diborane
dose at $T_{\mathrm{A}}=850$$\,^{\circ}$C. The legend for graphs
(b) and (c) is given in (c). (d) Longitudinal $\rho_{\square}$ and
transverse $R_{\mathrm{H}}$ magnetoresistance of the 3\,L sample,
annealed in-situ at $T_{\mathrm{A}}=350\,^{\circ}$C for 1\,min.}
\end{figure}

We use e-beam lithography to pattern Hall bars from the $\delta$-layer
samples and anneal them at different temperatures $T_{\mathrm{A}}$
to study activation of the dopants. Figure \ref{fig:2}a shows the
resistivity per square $\rho_{\square}$ as a function of measurement
temperature $T_{\mathrm{M}}$ for a $\delta$-layer sample with a
dose of 570\,L annealed for 2\,min in forming gas.
For $T_{\mathrm{M}}>35$\,K, substrate conductivity strongly influences
the measured resistivity. Below this temperature the carriers in the
low-doped substrate freeze out and measurements reflect the properties
of the boron $\delta$-layers. In the following, only this lower
temperature range is considered. $\rho_{\square}$ clearly drops with
increasing $T_{\mathrm{A}}$, giving values as low as 300$\;\Omega/\square$
for anneals at 850$\,^{\circ}$C. For $T_{\mathrm{A}}>250\,^{\circ}$C
the resistivities of the $\delta$-layers show only a weak temperature
dependence down to $T_{\mathrm{M}}=0.1$\,K (not shown).

Figure \ref{fig:2}b shows the sheet density $n_{\mathrm{S}}$ and
Figure \ref{fig:2}c the Hall mobility $\mu$ of all the samples that
were investigated. Without additional annealing we find a sheet density
of $10^{14}\mathrm{cm}^{-2}$ for all samples and little changes up
to $T_{\mathrm{A}}=$ 700$\,^{\circ}$C. From the bulk diffusivity
of boron in silicon we estimate that the full width at half the maximum
of the boron distribution starts to increase at $T_{\mathrm{A}}=$
650$\,^{\circ}$C, as shown by the dashed line in Figure \ref{fig:2}b
(calculated for a $2\;$min anneal). We find that $n_{\mathrm{S}}$
increases in sync with this broadening of the boron distribution,
reaching densities as high as $4\times10^{14}\mathrm{cm}^{-2}$. This
corresponds to the maximum density expected from the surface reaction
model put forward in Ref.~\cite{wang1996a} where $\mathrm{BH_{2}}$
and $\mathrm{H}$ terminate the broken $\pi$-bond of every surface
Si dimer. The observed increase in active carrier density can be explained
by two possible dopant (de)activation processes. First, the low temperature
overgrowth may leave the dopant atoms in interstitial sites or generally
create crystal defects at the growth interface that deactivate a fraction
of the charge carriers. However, from previous experiments with phosphorus
we do not expect this to be a significant concern. The second effect
is dopant deactivation because of the proximity of the boron atoms.
The broadening of the $\delta$-layer with increasing $T_{\mathrm{A}}$
leads to an increased spacing between dopants and a decrease in local
density $\rho_{3D}$. For $T_{\mathrm{A}}=750$$\,^{\circ}$C, we
estimate $\rho_{3D}=2\times10^{21}\mathrm{cm}^{-3}$. At this density
theoretical considerations indicate that, depending on the microscopic
arrangement, only about one third of the dopants are expected to be
active~\cite{03LuoAA}. This is in agreement with our observations.
However, it would mean that only about one boron atom per cubic Si
unit cell can be active, which is significantly less than what was
estimated in samples created by other techniques~\cite{06BustarretAA,weir1994}.
Furthermore, it is not entirely clear why the dosing procedure we
use always gives roughly the same initial density of $10^{14}cm^{-2}$
irrespective of the actual exposure dose. Contrary to what one may
have expected from previous STM studies~\cite{wang1996a,yu1986},
we find no dependence on exposure dose for doses of 90\,L or higher
after annealing at 850$\,^{\circ}$C (see inset in Figure \ref{fig:2}b).
The only sample which shows a sheet density that is clearly limited
by the initial exposure dose is the 3\,L sample (filled stars). 

The mobility $\mu$ is expected to be determined by ionised impurity
scattering. For this situation it was discussed in Ref.~\cite{weir1994}
that $\mu$ should depend only on the active carrier concentration,
similar to findings with n-type dopants. However, we find that $\mu$
increases in a more gradual fashion with $T_{\mathrm{A}}$ than the
sheet density. This may indicate that disorder at the regrowth interface
or clustering of the dopants within the $\delta$-layer also play
a role for $\mu$. The fact that the 3\,L sample shows a slightly
higher mobility for low $T_{\mathrm{A}}$ supports this, since we
expect the initial dopant separation to be larger in this situation.
Moreover, this sample has a factor of three lower density than the
other samples at $T_{\mathrm{A}}=850$$\,^{\circ}$C, while the mobility
is again similar. In an attempt to fully activate the dopants while
limiting the broadening of the $\delta$-layer, we also performed short
13\,sec in-situ anneals using direct current heating of two highly
dosed samples at $T_{\mathrm{A}}=850$$\,^{\circ}$C. We anneal the
samples after overgrowth with 2\,nm of silicon and before standard
encapsulation with silicon. This is indicated by the two cross symbols
in Figures \ref{fig:2}b, c. We find that we can fully activate the
dopants even though the FWHM of the dopant distribution is estimated
to be only 1.4\,nm. The mobility is, however, clearly reduced which
is consistent with the expected higher value of $\rho_{3D}$. Note
that it is difficult
to control the short duration and temperature of the anneal in-situ and we therefore expect more variability of the measured parameters for these samples.

This series of experiments shows that for full activation of the boron
acceptors annealing temperatures $T_{\mathrm{A}}>800\,^{\circ}$C
are necessary. Nevertheless, it also demonstrates that boron can be
incorporated at temperatures below $450\,^{\circ}$C, which is both
compatible with hydrogen resist lithography and gives resistivity
values $\rho_{\square}$ similar to those in n-type devices~\cite{05SimmonsAA}.
To support this, Figure \ref{fig:2}d shows the longitudinal magnetoresistance
of the 3\,L boron sample after annealing at $350\,^{\circ}$C for
1\,min, i.e. using identical conditions as for phosphorus donor device
fabrication. In contrast to phosphorus $\delta$-layers a positive
magnetoresistance is expected for boron~\cite{dai1992} and we find
a roughly four times higher $\rho_{\square}$, which is in equal part
due to a lower $\mu$ and $n_{\mathrm{S}}$.

\subsection{p-type nanowire and gap}

In the following we combine boron p-type $\delta$-doping
with STM lithography and create
a p-type device with three terminals containing a nanowire on one side and a gap on the other. The temperatures and times used
for the different steps of device fabrication are the same as in \ref{fig_SIMS}a.
The detailed description of the device fabrication process and ebeam
contacting procedure can be found in \cite{Pascher2016}. 

\begin{figure}[t!]
\centering{}\includegraphics{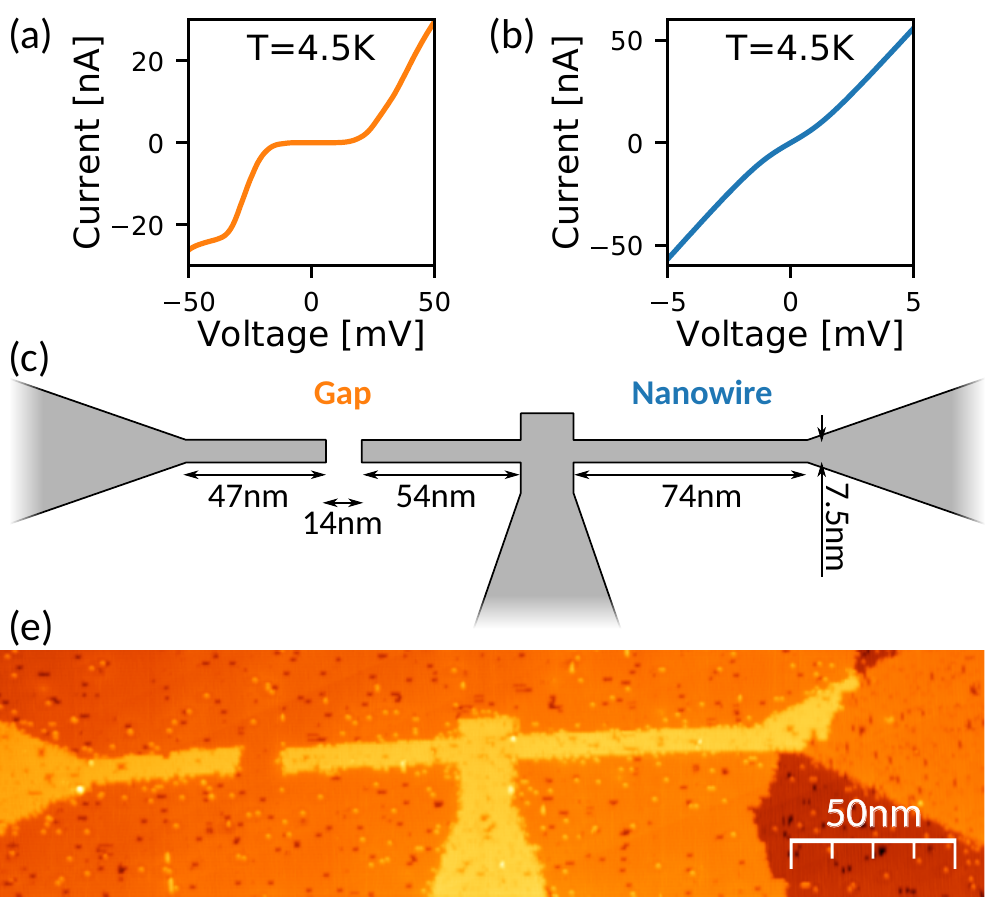}\caption{\label{fig_p_NW}Electrical measurements of the STM defined boron
doped dopant device. (a, b) Current-voltage characteristics of the
gap and nanowire section of the device. (c) Schematic drawing of the
device with dimensions and (d) the STM image of the device area after
the desorption of the H layer with the STM. }
\end{figure}

The device outline with the patterned dimensions is shown in Figure \ref{fig_p_NW}c along with an STM image of the desorbed pattern before dosing with diborane in Figure \ref{fig_p_NW}d. Figures \ref{fig_p_NW}a,
b show the electrical measurements of the Gap and Nanowire segment
at a temperature of 2.5\,K. The nanowire
shows Ohmic conductivity around 0\,V bias while the gap only shows
current after reaching a threshold voltage of about 20\,mV, as expected
for a tunnel junction. This threshold of 20\,mV is slightly
lower than the voltage observed for a tunnel junction with a similar
geometry but for a phosphorus doped device \cite{Pascher2016}. By
measuring the resistance of the nanowire section we can estimate the
sheet resistivity of the boron doped layer to be 5\,$\mathrm{k\Omega\cdot\square^{-1}}$. 

These results show that the p-type nanowire device behaves in a similar
way as expected from a corresponding n-type dopant device.
The sheet resistivity of the p-type layer is slightly higher due to limited dopant activation at the temperatures available
for this process. Moreover, the sheet resistivity as well as contact
resistance has been more variable for p-type devices than for
corresponding phosphorus doped n-type devices. The critical step is
probably the activation anneal which would have to be very tightly
controlled in order to achieve precisely reproducible resistivities. 

\subsection{pn-junction}

After developing the method for p-type $\delta$-doping and constructing a p-type
nanoscale device, an obvious next step is to combine p- and
n-type doping in a single sample to fabricate a pn-device. The most
simple bipolar device is a two terminal pn-junction. 

\begin{figure}[t!]
\centering{}\includegraphics{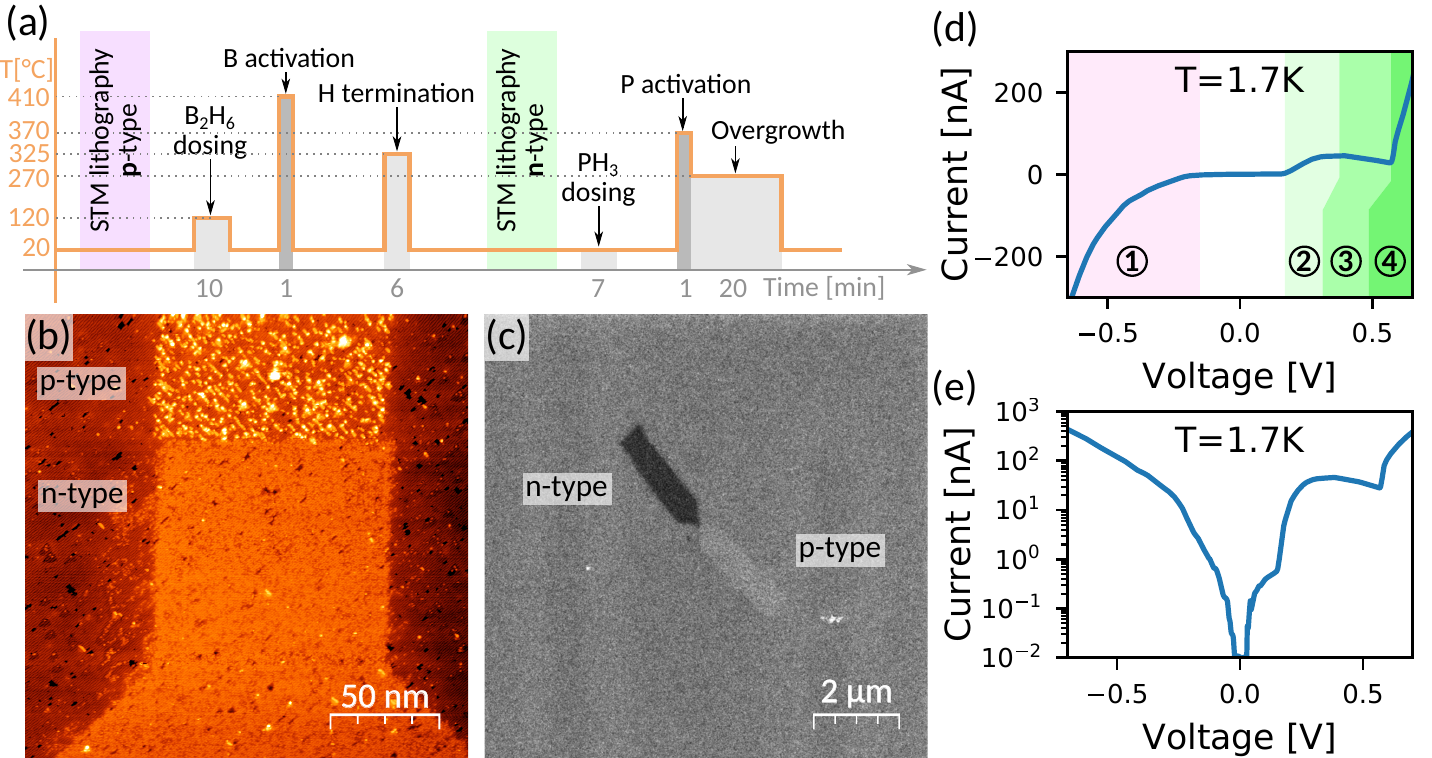}\caption{\label{fig_pn}Fabrication of a bipolar dopant device with STM lithogrpahy.
(a) Preparation procedure, (b) STM image of the central part of the
pn junction right after desorption of the area for phosphorus doping
(lower part). In the upper section, boron doped and incorporated area
is visible. (c) Ex-situ SEM image of the pn-device from (c). Linear
(d) and semilog plot of the current-voltage characteristics of the
pn junction. }
\end{figure}

The procedure for bipolar device fabrication is a two step process
which is outlined in Figure \ref{fig_pn}a. In a first step, we use the STM to define the areas for p-type doping, dose the sample with
diborane and perform the activation anneal. The reason we start with
p-type doping is that the thermal budget is slightly higher in
this case and, as such, we prevent unnecessary exposure of the
n-type doped areas to these higher temperatures. The temperature we
use for the boron activation (\SI{410}{\celsius}) already leads to
a partial desorption of the H passivation layer from the Si surface
and renewal of the H layer is required by exposing the sample to atomic hydrogen at a temperature of \SI{323}{\celsius}. Subsequently,
the sample is transferred back to the STM stage and hydrogen resist patterning for the n-type doping step is performed. The critical step here is alignment of the n-type pattern to the previously doped p-type area. This is achieved by
careful placement of the device within the optical localization marker
pattern using an optical microscope. We then use the STM to scan a $8\times8$ micron area and identify the patch with ejected silicon from the first doping step. Figure \ref{fig_pn}b shows
an STM image of the central part of the pn-junction after the desorption
of the area for n-type doping in the lower part of the pattern. The boron
doped area can be seen in the upper part of the image. After
STM desorption, the sample is dosed with phosphine, the activation
anneal is performed and, finally, the sample is overgrown with a 20\,nm
layer of intrinsic Si. Scanning electron (SEM) image of the finished
pn-device is presented in Figure \ref{fig_pn}c. The n- (p-type) areas
show dark (bright) contrast in SEM imaging with the in-lens detector respectively. This behavior is often observed when imaging doped semiconductors and stems from the energy sensitivity of this detector. Subsequently, the device was contacted using e-beam-defined metallic contacts and the sample was placed in a package and wire bonded. 

Current-voltage characteristic of the pn-junction measured at a temperature of 1.7\,K are shown in Figures \ref{fig_pn}d,
e. The pn-junction is expected to exhibit assymetric behavior with
higher conductivity in the forward direction (positive voltage). This
behavior can be seen for voltages higher than $\sim$0.6\,V (region
4 in Figure \ref{fig_pn}d). Further, for the reverse-bias direction (negative voltage) we can see a gradual increase of the current (region 1). This is related to the high doping density on both sides of the pn junction
- the depletion region around the pn junction is narrow and allows
tunneling of the charge carriers. The conduction is also dominated
by the tunneling current in the forward direction in region 2. In
region 3, the current again drops which is caused by misalignment
of the bands in p- and n-doped regions. This region of negative differential
resistance is a characteristic feature for highly doped pn junctions
(Esaki diode). 

Fabrication of bipolar devices is not substantially more demanding
than fabrication of an n- or p-type dopant device. The only additional
step is the alignment of the STM tip to the previously doped region
and this can be solved by proper choice of device position with respect
to the optical localization markers. However, similarly to the remark
about p-type devices in the previous section, there is larger variability
in the sheet resistivities and contact resistances. For a number of
pn-devices we even encountered issues with non-functional contacts,
mostly to the p-type part of the device. We believe that these problems
can be solved by better control of the boron activation step, however,
these issues will require more detailed investigation before the technique
can be adapted as routine method for bipolar dopant device fabrication. 

\section{Conclusions}

In conclusion, we have shown that hydrogen resist lithography can
be used for STM-patterning of p-type nanostructures with a resolution better
than $2\,$nm. Even though full activation of the dopants can only
be reached at $T_{\mathrm{A}}=$ 850$\,^{\circ}$C , we demonstrate
that boron $\delta$-layers with resistivities below $3.3\;$k$\Omega$
can be fabricated using dosing and incorporation temperatures that
are compatible with hydrogen resist lithography. To demonstrate the
feasibility of this approach we fabricated a 7.5\,nm wide p-type
dopant nanowire which shows ohmic conductivity and resistivity of
about 5\,$\mathrm{k\Omega\square^{-1}}$. By combining the p-type
doping with estalished n-type doping we also prepared the first bipolar
dopant device - a 100\,nm wide pn-junction which behaves similar
to an Esaki diode. These results open the way for fabrication of
a large variety of more complex p-type and bipolar nanoscale dopant
devices. 

\section{Methods}

Samples are fabricated from $2.7\times9$\,mm, n-type Si(001) wafer
pieces with a resistivity of 0.1-1.0 $\Omega\,$cm. For sample preparation
and hydrogen termination we follow the procedure outlined in Ref.~\cite{09FuhrerAA}. 

For desorption of hydrogen with the STM tip we typically used a tunneling
current setpoint of 3\,nA and sample bias of +4.1\,V (for high resolution
patterning of nanoscale parts) up to 6.5\,V (for fast low resolution
desorption of large areas). 

Dopant pre-cursors are introduced through a leak valve directly into
the STM chamber containing an Omicron VT STM/AFM system. For diborane
dosing we use 30\% diborane in molecular hydrogen (H$_{2}$). The
sample is positioned directly in the gas flux while maintaining the
pressure in the chamber constant. The doses given refer to total
gas exposure as calculated from chamber pressure multiplied by exposure
time. This means that effective diborane doses are about a factor
of three lower than what we state.

Silicon overgrowth is performed by sublimation from an intrinsic silicon
filament with a typical rate of 0.4~nm/min while keeping the sample
at 270$\,^{\circ}$C. After this, further processing occurs ex-situ
using standard lithography techniques. For contacting the p-type devices
we used Pt contacts, since the Fermi level in Pt is closer to that
of degenerately doped p-type Si leading to a lower Schottky barrier.
In the case of bipolar devices, we used Al contacts. 

The SIMS profiles (Cameca SC Ultra) have been acquired using a $250\,$eV $O_2^+$ primary beam with an incidence angle of 46$^{\circ}$ to the sample normal (calculated based on the potentials).

The boron concentration was quantified by measuring a boron reference sample with a known boron concentration and by applying the RSF method point-by-point relative to the Si matrix signal. The total crater depth of the reference sample was measured by stylus profilometry allowing to determine the current-normalized sputter rate value. This value was then applied point-by-point to all profiles relative to their primary current curves to account for sputter rate variations with the primary beam current.

The authors gratefully acknowledge financial support from EU-FET grants
SiAM 610637, PAMS 610446 and from the Swiss NCCR QSIT. 

\bibliographystyle{apsrev}

\begin{thebibliography}{25}
\expandafter\ifx\csname natexlab\endcsname\relax\def\natexlab#1{#1}\fi
\expandafter\ifx\csname bibnamefont\endcsname\relax
  \def\bibnamefont#1{#1}\fi
\expandafter\ifx\csname bibfnamefont\endcsname\relax
  \def\bibfnamefont#1{#1}\fi
\expandafter\ifx\csname citenamefont\endcsname\relax
  \def\citenamefont#1{#1}\fi
\expandafter\ifx\csname url\endcsname\relax
  \def\url#1{\texttt{#1}}\fi
\expandafter\ifx\csname urlprefix\endcsname\relax\def\urlprefix{URL }\fi
\providecommand{\bibinfo}[2]{#2}
\providecommand{\eprint}[2][]{\url{#2}}

\bibitem[{\citenamefont{Frank et~al.}(2001)\citenamefont{Frank, Dennard, Nowak,
  Solomon, Taur, and Wong}}]{01FrankAA}
\bibinfo{author}{\bibfnamefont{D.~J.} \bibnamefont{Frank}},
  \bibinfo{author}{\bibfnamefont{R.~H.} \bibnamefont{Dennard}},
  \bibinfo{author}{\bibfnamefont{E.}~\bibnamefont{Nowak}},
  \bibinfo{author}{\bibfnamefont{P.~M.} \bibnamefont{Solomon}},
  \bibinfo{author}{\bibfnamefont{Y.}~\bibnamefont{Taur}}, \bibnamefont{and}
  \bibinfo{author}{\bibfnamefont{H.-S.~P.} \bibnamefont{Wong}},
  \bibinfo{journal}{Proceedings of the IEEE} \textbf{\bibinfo{volume}{89}},
  \bibinfo{pages}{259} (\bibinfo{year}{2001}).

\bibitem[{\citenamefont{Ho et~al.}(2008)\citenamefont{Ho, Yerushalmi, Jacobson,
  Fan, Alley, and Javey}}]{08HoAA}
\bibinfo{author}{\bibfnamefont{J.~C.} \bibnamefont{Ho}},
  \bibinfo{author}{\bibfnamefont{R.}~\bibnamefont{Yerushalmi}},
  \bibinfo{author}{\bibfnamefont{Z.~A.} \bibnamefont{Jacobson}},
  \bibinfo{author}{\bibfnamefont{Z.}~\bibnamefont{Fan}},
  \bibinfo{author}{\bibfnamefont{R.~L.} \bibnamefont{Alley}}, \bibnamefont{and}
  \bibinfo{author}{\bibfnamefont{A.}~\bibnamefont{Javey}},
  \bibinfo{journal}{Nature Materials} \textbf{\bibinfo{volume}{7}},
  \bibinfo{pages}{62} (\bibinfo{year}{2008}).

\bibitem[{\citenamefont{Colombeau et~al.}(2014)\citenamefont{Colombeau, Guo,
  Gossman, Khaja, Pradhan, Waite, Thomidis, Shim, Henry, and
  Variam}}]{14ColombeauAA}
\bibinfo{author}{\bibfnamefont{B.}~\bibnamefont{Colombeau}},
  \bibinfo{author}{\bibfnamefont{B.}~\bibnamefont{Guo}},
  \bibinfo{author}{\bibfnamefont{H.-J.} \bibnamefont{Gossman}},
  \bibinfo{author}{\bibfnamefont{F.}~\bibnamefont{Khaja}},
  \bibinfo{author}{\bibfnamefont{N.}~\bibnamefont{Pradhan}},
  \bibinfo{author}{\bibfnamefont{K.~V.} \bibnamefont{Waite},
  \bibfnamefont{A.~adn~Rao}},
  \bibinfo{author}{\bibfnamefont{C.}~\bibnamefont{Thomidis}},
  \bibinfo{author}{\bibfnamefont{K.-H.} \bibnamefont{Shim}},
  \bibinfo{author}{\bibfnamefont{T.}~\bibnamefont{Henry}}, \bibnamefont{and}
  \bibinfo{author}{\bibfnamefont{N.}~\bibnamefont{Variam}},
  \bibinfo{journal}{Phys. Stat. Sol. (a)} \textbf{\bibinfo{volume}{211}},
  \bibinfo{pages}{101} (\bibinfo{year}{2014}).

\bibitem[{\citenamefont{Duffy et~al.}(2014)\citenamefont{Duffy, Shayesteh,
  Thomas, Pelucchi, Yu, Gangnaik, Georiev, carolan, Petkov, Long
  et~al.}}]{14DuffyAA}
\bibinfo{author}{\bibfnamefont{R.}~\bibnamefont{Duffy}},
  \bibinfo{author}{\bibfnamefont{M.}~\bibnamefont{Shayesteh}},
  \bibinfo{author}{\bibfnamefont{K.}~\bibnamefont{Thomas}},
  \bibinfo{author}{\bibfnamefont{E.}~\bibnamefont{Pelucchi}},
  \bibinfo{author}{\bibfnamefont{R.}~\bibnamefont{Yu}},
  \bibinfo{author}{\bibfnamefont{A.}~\bibnamefont{Gangnaik}},
  \bibinfo{author}{\bibfnamefont{Y.~M.} \bibnamefont{Georiev}},
  \bibinfo{author}{\bibfnamefont{P.}~\bibnamefont{carolan}},
  \bibinfo{author}{\bibfnamefont{N.}~\bibnamefont{Petkov}},
  \bibinfo{author}{\bibfnamefont{B.}~\bibnamefont{Long}}, \bibnamefont{et~al.},
  \bibinfo{journal}{J. Mat. Chem. C} \textbf{\bibinfo{volume}{2}},
  \bibinfo{pages}{9248} (\bibinfo{year}{2014}).

\bibitem[{\citenamefont{Muhonen et~al.}(2014)\citenamefont{Muhonen, Debollain,
  Laucht, Hudson, Sekiguchi, Itoh, Jamison, McCallum, Dzurak, and
  Morello}}]{14MuhonenAA}
\bibinfo{author}{\bibfnamefont{J.~T.} \bibnamefont{Muhonen}},
  \bibinfo{author}{\bibfnamefont{J.~P.} \bibnamefont{Debollain}},
  \bibinfo{author}{\bibfnamefont{A.}~\bibnamefont{Laucht}},
  \bibinfo{author}{\bibfnamefont{F.~E.} \bibnamefont{Hudson}},
  \bibinfo{author}{\bibfnamefont{T.}~\bibnamefont{Sekiguchi}},
  \bibinfo{author}{\bibfnamefont{K.~M.} \bibnamefont{Itoh}},
  \bibinfo{author}{\bibfnamefont{D.~N.} \bibnamefont{Jamison}},
  \bibinfo{author}{\bibfnamefont{J.~C.} \bibnamefont{McCallum}},
  \bibinfo{author}{\bibfnamefont{A.~S.} \bibnamefont{Dzurak}},
  \bibnamefont{and} \bibinfo{author}{\bibfnamefont{A.}~\bibnamefont{Morello}},
  \bibinfo{journal}{Nature Nanotech.} \textbf{\bibinfo{volume}{9}},
  \bibinfo{pages}{986} (\bibinfo{year}{2014}).

\bibitem[{\citenamefont{Roche et~al.}(2013)\citenamefont{Roche, Riwar, Voisin,
  Dupont-Ferrier, Wacquez, Vinet, Sanquer, Splettstoesser, and
  Jehl}}]{13RocheAA}
\bibinfo{author}{\bibfnamefont{B.}~\bibnamefont{Roche}},
  \bibinfo{author}{\bibfnamefont{R.-P.} \bibnamefont{Riwar}},
  \bibinfo{author}{\bibfnamefont{B.}~\bibnamefont{Voisin}},
  \bibinfo{author}{\bibfnamefont{E.}~\bibnamefont{Dupont-Ferrier}},
  \bibinfo{author}{\bibfnamefont{R.}~\bibnamefont{Wacquez}},
  \bibinfo{author}{\bibfnamefont{M.}~\bibnamefont{Vinet}},
  \bibinfo{author}{\bibfnamefont{M.}~\bibnamefont{Sanquer}},
  \bibinfo{author}{\bibfnamefont{J.}~\bibnamefont{Splettstoesser}},
  \bibnamefont{and} \bibinfo{author}{\bibfnamefont{X.}~\bibnamefont{Jehl}},
  \bibinfo{journal}{Nature Communications} \textbf{\bibinfo{volume}{4}},
  \bibinfo{pages}{1581} (\bibinfo{year}{2013}).

\bibitem[{\citenamefont{Schofield et~al.}(2003)\citenamefont{Schofield, Curson,
  Simmons, Rue{\ss}, Hallam, Oberbeck, and Clark}}]{03SchofieldAA}
\bibinfo{author}{\bibfnamefont{S.~R.} \bibnamefont{Schofield}},
  \bibinfo{author}{\bibfnamefont{N.~J.} \bibnamefont{Curson}},
  \bibinfo{author}{\bibfnamefont{M.~Y.} \bibnamefont{Simmons}},
  \bibinfo{author}{\bibfnamefont{F.~J.} \bibnamefont{Rue{\ss}}},
  \bibinfo{author}{\bibfnamefont{T.}~\bibnamefont{Hallam}},
  \bibinfo{author}{\bibfnamefont{L.}~\bibnamefont{Oberbeck}}, \bibnamefont{and}
  \bibinfo{author}{\bibfnamefont{R.~G.} \bibnamefont{Clark}},
  \bibinfo{journal}{Phys. Rev. Lett.} \textbf{\bibinfo{volume}{91}},
  \bibinfo{pages}{136104} (\bibinfo{year}{2003}).

\bibitem[{\citenamefont{Fuechsle et~al.}(2012)\citenamefont{Fuechsle, Miwa,
  Mahapatra, Ryu, Lee, Warschkow, Hollenberg, Klimeck, and
  Simmons}}]{12FuechsleAA}
\bibinfo{author}{\bibfnamefont{M.}~\bibnamefont{Fuechsle}},
  \bibinfo{author}{\bibfnamefont{J.~A.} \bibnamefont{Miwa}},
  \bibinfo{author}{\bibfnamefont{S.}~\bibnamefont{Mahapatra}},
  \bibinfo{author}{\bibfnamefont{H.}~\bibnamefont{Ryu}},
  \bibinfo{author}{\bibfnamefont{S.}~\bibnamefont{Lee}},
  \bibinfo{author}{\bibfnamefont{O.}~\bibnamefont{Warschkow}},
  \bibinfo{author}{\bibfnamefont{L.~C.~L.} \bibnamefont{Hollenberg}},
  \bibinfo{author}{\bibfnamefont{G.}~\bibnamefont{Klimeck}}, \bibnamefont{and}
  \bibinfo{author}{\bibfnamefont{M.~Y.} \bibnamefont{Simmons}},
  \bibinfo{journal}{Nature Nanotech.} \textbf{\bibinfo{volume}{7}},
  \bibinfo{pages}{242} (\bibinfo{year}{2012}).

\bibitem[{\citenamefont{Simmons et~al.}(2005)\citenamefont{Simmons, Ruess, Goh,
  Hallam, Schofield, Oberbeck, Curson, Hamilton, Butcher, Clark
  et~al.}}]{05SimmonsAA}
\bibinfo{author}{\bibfnamefont{M.}~\bibnamefont{Simmons}},
  \bibinfo{author}{\bibfnamefont{F.}~\bibnamefont{Ruess}},
  \bibinfo{author}{\bibfnamefont{K.}~\bibnamefont{Goh}},
  \bibinfo{author}{\bibfnamefont{T.}~\bibnamefont{Hallam}},
  \bibinfo{author}{\bibfnamefont{S.}~\bibnamefont{Schofield}},
  \bibinfo{author}{\bibfnamefont{L.}~\bibnamefont{Oberbeck}},
  \bibinfo{author}{\bibfnamefont{N.}~\bibnamefont{Curson}},
  \bibinfo{author}{\bibfnamefont{A.}~\bibnamefont{Hamilton}},
  \bibinfo{author}{\bibfnamefont{M.}~\bibnamefont{Butcher}},
  \bibinfo{author}{\bibfnamefont{R.}~\bibnamefont{Clark}},
  \bibnamefont{et~al.}, \bibinfo{journal}{Molecular Simulation}
  \textbf{\bibinfo{volume}{31}}, \bibinfo{pages}{505} (\bibinfo{year}{2005}).

\bibitem[{\citenamefont{van~der Heijden et~al.}(2015)\citenamefont{van~der
  Heijden, Salfi, Mol, Verduijn, Tettamanzi, Hamilton, Collaert, and
  Rogge}}]{15HeijdenAA}
\bibinfo{author}{\bibfnamefont{J.}~\bibnamefont{van~der Heijden}},
  \bibinfo{author}{\bibfnamefont{J.}~\bibnamefont{Salfi}},
  \bibinfo{author}{\bibfnamefont{J.~A.} \bibnamefont{Mol}},
  \bibinfo{author}{\bibfnamefont{J.}~\bibnamefont{Verduijn}},
  \bibinfo{author}{\bibfnamefont{G.~C.} \bibnamefont{Tettamanzi}},
  \bibinfo{author}{\bibfnamefont{A.~R.} \bibnamefont{Hamilton}},
  \bibinfo{author}{\bibfnamefont{N.}~\bibnamefont{Collaert}}, \bibnamefont{and}
  \bibinfo{author}{\bibfnamefont{S.}~\bibnamefont{Rogge}},
  \bibinfo{journal}{Nano Letters} \textbf{\bibinfo{volume}{14}},
  \bibinfo{pages}{1492} (\bibinfo{year}{2015}).

\bibitem[{\citenamefont{Wang et~al.}(1996)\citenamefont{Wang, Shan, and
  Hamers}}]{wang1996a}
\bibinfo{author}{\bibfnamefont{Y.}~\bibnamefont{Wang}},
  \bibinfo{author}{\bibfnamefont{J.}~\bibnamefont{Shan}}, \bibnamefont{and}
  \bibinfo{author}{\bibfnamefont{R.~J.} \bibnamefont{Hamers}},
  \bibinfo{journal}{J. Vac. Sci. Techn. B} \textbf{\bibinfo{volume}{14}},
  \bibinfo{pages}{1038} (\bibinfo{year}{1996}).

\bibitem[{\citenamefont{Yu et~al.}(1986)\citenamefont{Yu, Vitkavage, and
  Meyerson}}]{yu1986}
\bibinfo{author}{\bibfnamefont{M.~L.} \bibnamefont{Yu}},
  \bibinfo{author}{\bibfnamefont{D.~J.} \bibnamefont{Vitkavage}},
  \bibnamefont{and} \bibinfo{author}{\bibfnamefont{B.~S.}
  \bibnamefont{Meyerson}}, \bibinfo{journal}{J. Appl. Phys.}
  \textbf{\bibinfo{volume}{59}}, \bibinfo{pages}{4032} (\bibinfo{year}{1986}).

\bibitem[{\citenamefont{Oura et~al.}(1990)\citenamefont{Oura, Yamane, Umezawa,
  Naitoh, Shoji, and Hanawa}}]{90OuraAA}
\bibinfo{author}{\bibfnamefont{K.}~\bibnamefont{Oura}},
  \bibinfo{author}{\bibfnamefont{J.}~\bibnamefont{Yamane}},
  \bibinfo{author}{\bibfnamefont{K.}~\bibnamefont{Umezawa}},
  \bibinfo{author}{\bibfnamefont{M.}~\bibnamefont{Naitoh}},
  \bibinfo{author}{\bibfnamefont{F.}~\bibnamefont{Shoji}}, \bibnamefont{and}
  \bibinfo{author}{\bibfnamefont{T.}~\bibnamefont{Hanawa}},
  \bibinfo{journal}{Phys. Rev. B} \textbf{\bibinfo{volume}{41}},
  \bibinfo{pages}{1200} (\bibinfo{year}{1990}).

\bibitem[{\citenamefont{D{\"u}rr et~al.}(2002)\citenamefont{D{\"u}rr, Hu,
  Biedermann, H{\"o}fer, and Heinz}}]{durr2002}
\bibinfo{author}{\bibfnamefont{M.}~\bibnamefont{D{\"u}rr}},
  \bibinfo{author}{\bibfnamefont{Z.}~\bibnamefont{Hu}},
  \bibinfo{author}{\bibfnamefont{A.}~\bibnamefont{Biedermann}},
  \bibinfo{author}{\bibfnamefont{U.}~\bibnamefont{H{\"o}fer}},
  \bibnamefont{and} \bibinfo{author}{\bibfnamefont{T.~F.} \bibnamefont{Heinz}},
  \bibinfo{journal}{Phys. Rev. Lett.} \textbf{\bibinfo{volume}{88}},
  \bibinfo{pages}{046104} (\bibinfo{year}{2002}).

\bibitem[{\citenamefont{Wang and Hamers}(1995)}]{wang1995}
\bibinfo{author}{\bibfnamefont{Y.}~\bibnamefont{Wang}} \bibnamefont{and}
  \bibinfo{author}{\bibfnamefont{R.~J.} \bibnamefont{Hamers}},
  \bibinfo{journal}{J. Vac. Sci. Technol. A} \textbf{\bibinfo{volume}{13}},
  \bibinfo{pages}{1431} (\bibinfo{year}{1995}).

\bibitem[{\citenamefont{Wilson et~al.}(2006)\citenamefont{Wilson, Warschkow,
  Marks, Curson, Schofield, Reusch, Radny, Smith, McKenzie, and
  Simmons}}]{06WilsonAA}
\bibinfo{author}{\bibfnamefont{H.~F.} \bibnamefont{Wilson}},
  \bibinfo{author}{\bibfnamefont{O.}~\bibnamefont{Warschkow}},
  \bibinfo{author}{\bibfnamefont{N.~A.} \bibnamefont{Marks}},
  \bibinfo{author}{\bibfnamefont{N.~J.} \bibnamefont{Curson}},
  \bibinfo{author}{\bibfnamefont{S.~R.} \bibnamefont{Schofield}},
  \bibinfo{author}{\bibfnamefont{T.~C.~G.} \bibnamefont{Reusch}},
  \bibinfo{author}{\bibfnamefont{M.~W.} \bibnamefont{Radny}},
  \bibinfo{author}{\bibfnamefont{P.~V.} \bibnamefont{Smith}},
  \bibinfo{author}{\bibfnamefont{D.~R.} \bibnamefont{McKenzie}},
  \bibnamefont{and} \bibinfo{author}{\bibfnamefont{M.~Y.}
  \bibnamefont{Simmons}}, \bibinfo{journal}{Phys. Rev. B}
  \textbf{\bibinfo{volume}{74}}, \bibinfo{pages}{195310}
  (\bibinfo{year}{2006}).

\bibitem[{\citenamefont{Liu et~al.}(2008)\citenamefont{Liu, Zhang, and
  Zhu}}]{liu2008}
\bibinfo{author}{\bibfnamefont{Z.}~\bibnamefont{Liu}},
  \bibinfo{author}{\bibfnamefont{Z.}~\bibnamefont{Zhang}}, \bibnamefont{and}
  \bibinfo{author}{\bibfnamefont{X.}~\bibnamefont{Zhu}},
  \bibinfo{journal}{Phys. Rev. B} \textbf{\bibinfo{volume}{77}},
  \bibinfo{pages}{035322} (\bibinfo{year}{2008}).

\bibitem[{\citenamefont{Goh et~al.}(2004)\citenamefont{Goh, Oberbeck, Simmons,
  Hamilton, and Clark}}]{Goh2004}
\bibinfo{author}{\bibfnamefont{K.~E.~J.} \bibnamefont{Goh}},
  \bibinfo{author}{\bibfnamefont{L.}~\bibnamefont{Oberbeck}},
  \bibinfo{author}{\bibfnamefont{M.~Y.} \bibnamefont{Simmons}},
  \bibinfo{author}{\bibfnamefont{A.~R.} \bibnamefont{Hamilton}},
  \bibnamefont{and} \bibinfo{author}{\bibfnamefont{R.~G.} \bibnamefont{Clark}},
  \bibinfo{journal}{Applied Physics Letters} \textbf{\bibinfo{volume}{85}},
  \bibinfo{pages}{4953} (\bibinfo{year}{2004}), ISSN \bibinfo{issn}{00036951}.

\bibitem[{\citenamefont{Keizer et~al.}(2015)\citenamefont{Keizer, Koelling,
  Koenraad, and Simmons}}]{Keizer2015}
\bibinfo{author}{\bibfnamefont{J.~G.} \bibnamefont{Keizer}},
  \bibinfo{author}{\bibfnamefont{S.}~\bibnamefont{Koelling}},
  \bibinfo{author}{\bibfnamefont{P.~M.} \bibnamefont{Koenraad}},
  \bibnamefont{and} \bibinfo{author}{\bibfnamefont{M.~Y.}
  \bibnamefont{Simmons}}, \bibinfo{journal}{ACS Nano}
  \textbf{\bibinfo{volume}{9}}, \bibinfo{pages}{12537} (\bibinfo{year}{2015}),
  ISSN \bibinfo{issn}{1936086X}.

\bibitem[{\citenamefont{Luo et~al.}(2003)\citenamefont{Luo, Zhang, and
  Wei}}]{03LuoAA}
\bibinfo{author}{\bibfnamefont{X.}~\bibnamefont{Luo}},
  \bibinfo{author}{\bibfnamefont{S.~B.} \bibnamefont{Zhang}}, \bibnamefont{and}
  \bibinfo{author}{\bibfnamefont{S.-H.} \bibnamefont{Wei}},
  \bibinfo{journal}{Phys. Rev. Lett.} \textbf{\bibinfo{volume}{90}},
  \bibinfo{pages}{026103} (\bibinfo{year}{2003}).

\bibitem[{\citenamefont{Bustarret et~al.}(2006)\citenamefont{Bustarret,
  Marcenat, Achatz, Kacmarcik, L{\'e}vy, Huxley, Ort{\'e}ga, Bourgeois, Blase,
  D{\'e}barre et~al.}}]{06BustarretAA}
\bibinfo{author}{\bibfnamefont{E.}~\bibnamefont{Bustarret}},
  \bibinfo{author}{\bibfnamefont{C.}~\bibnamefont{Marcenat}},
  \bibinfo{author}{\bibfnamefont{P.}~\bibnamefont{Achatz}},
  \bibinfo{author}{\bibfnamefont{J.}~\bibnamefont{Kacmarcik}},
  \bibinfo{author}{\bibfnamefont{F.}~\bibnamefont{L{\'e}vy}},
  \bibinfo{author}{\bibfnamefont{A.}~\bibnamefont{Huxley}},
  \bibinfo{author}{\bibfnamefont{L.}~\bibnamefont{Ort{\'e}ga}},
  \bibinfo{author}{\bibfnamefont{E.}~\bibnamefont{Bourgeois}},
  \bibinfo{author}{\bibfnamefont{X.}~\bibnamefont{Blase}},
  \bibinfo{author}{\bibfnamefont{D.}~\bibnamefont{D{\'e}barre}},
  \bibnamefont{et~al.}, \bibinfo{journal}{Nature}
  \textbf{\bibinfo{volume}{444}}, \bibinfo{pages}{465} (\bibinfo{year}{2006}).

\bibitem[{\citenamefont{Weir et~al.}(1994)\citenamefont{Weir, Feldman, Monroe,
  , Grossmann, Headrick, and Hart}}]{weir1994}
\bibinfo{author}{\bibfnamefont{B.~E.} \bibnamefont{Weir}},
  \bibinfo{author}{\bibfnamefont{L.~C.} \bibnamefont{Feldman}},
  \bibinfo{author}{\bibfnamefont{D.}~\bibnamefont{Monroe}}, ,
  \bibinfo{author}{\bibfnamefont{H.-J.} \bibnamefont{Grossmann}},
  \bibinfo{author}{\bibfnamefont{R.~L.} \bibnamefont{Headrick}},
  \bibnamefont{and} \bibinfo{author}{\bibfnamefont{T.~R.} \bibnamefont{Hart}},
  \bibinfo{journal}{Appl. Phys. Lett.} \textbf{\bibinfo{volume}{65}},
  \bibinfo{pages}{737} (\bibinfo{year}{1994}).

\bibitem[{\citenamefont{Dai et~al.}(1992)\citenamefont{Dai, Zhang, and
  Sarachik}}]{dai1992}
\bibinfo{author}{\bibfnamefont{P.}~\bibnamefont{Dai}},
  \bibinfo{author}{\bibfnamefont{P.}~\bibnamefont{Zhang}}, \bibnamefont{and}
  \bibinfo{author}{\bibfnamefont{M.~P.} \bibnamefont{Sarachik}},
  \bibinfo{journal}{Phys. Rev. B} \textbf{\bibinfo{volume}{45}},
  \bibinfo{pages}{3984} (\bibinfo{year}{1992}).

\bibitem[{\citenamefont{Pascher et~al.}(2016)\citenamefont{Pascher, Hennel,
  Mueller, and Fuhrer}}]{Pascher2016}
\bibinfo{author}{\bibfnamefont{N.}~\bibnamefont{Pascher}},
  \bibinfo{author}{\bibfnamefont{S.}~\bibnamefont{Hennel}},
  \bibinfo{author}{\bibfnamefont{S.}~\bibnamefont{Mueller}}, \bibnamefont{and}
  \bibinfo{author}{\bibfnamefont{A.}~\bibnamefont{Fuhrer}},
  \bibinfo{journal}{New Journal of Physics} \textbf{\bibinfo{volume}{18}},
  \bibinfo{pages}{083001} (\bibinfo{year}{2016}).

\bibitem[{\citenamefont{Fuhrer et~al.}(2009)\citenamefont{Fuhrer, F{\"u}chsle,
  Reusch, Weber, and Simmons}}]{09FuhrerAA}
\bibinfo{author}{\bibfnamefont{A.}~\bibnamefont{Fuhrer}},
  \bibinfo{author}{\bibfnamefont{M.}~\bibnamefont{F{\"u}chsle}},
  \bibinfo{author}{\bibfnamefont{T.~C.~G.} \bibnamefont{Reusch}},
  \bibinfo{author}{\bibfnamefont{B.}~\bibnamefont{Weber}}, \bibnamefont{and}
  \bibinfo{author}{\bibfnamefont{M.~Y.} \bibnamefont{Simmons}},
  \bibinfo{journal}{Nano Letters} \textbf{\bibinfo{volume}{9}},
  \bibinfo{pages}{707} (\bibinfo{year}{2009}).

\end{thebibliography}

\end{document}